# THE ASTRONOMY OF TWO INDIAN TRIBES: THE BANJARAS AND THE KOLAMS

**M.N. Vahia**
*Tata Institute of Fundamental Research, Homi Bhabha Road, Mumbai 400 005, India.*
Email: vahia@tifr.res.in

**Ganesh Halkare**
*Indrayani Colony, Amravati, 444 607, India.*
Email: ganeshhalkare@gmail.com

**Kishore Menon**
*Tata Institute of Fundamental Research, Homi Bhabha Road, Mumbai 400 005, India.*
Email: kmenon@tifr.res.in

and

**Harini Calamur**
*1602 C Lloyds Estate, Vidvalankar, College Road, Wadala E., Mumbai 37, India.*
Email: calamur@gmail.com

**Abstract:** We report field studies of the astronomical beliefs of two Indian tribes – the Banjaras and the Kolams. The Banjaras are an ancient tribe connected with the gypsies of Europe while the Kolams have been foragers until recently. They share their landscape with each other and also with the Gonds whose astronomy was reported previously (Vahia and Halkare, 2013). The primary profession of the Banjaras was trade, based on the large-scale movement of goods over long distances, but their services were taken over by the railways about one hundred years ago. Since then the Banjaras have begun the long journey to a sedentary lifestyle. Meanwhile, the Kolams were foragers until about fifty years ago when the Government of India began to help them lead a settled life.

Here, we compare their astronomical beliefs of the Banjaras and the Kolams, which indicate the strong sense of identity that each community possesses. Our study also highlights their perspective about the sky and its relation to their daily lives. We show that apart from the absolute importance of the data on human perception of the sky, the data also reveal subtle aspects of interactions between physically co-located but otherwise isolated communities as well as their own lifestyles. We also show that there is a strong relationship between profession and perspective of the sky.

## 1 INTRODUCTION

India is one of the most complex regions in terms of co-habitation of a wide varieties of tribal societies that include the entire spectrum, from lifestyles akin to early hunter-gatherers to the most modern aspects of human civilisation. Many of these cultures are broadly classified as tribal, and they have remained intellectually isolated in an endogenous environment in spite of connections with all the trappings of modern civilisation. For example, they sell their goods at modern Indian markets and use other trappings of modern life, such as mobile phones very comfortably, but culturally and in daily life, they consciously keep themselves isolated from other communities and from urban people, and they try to preserve as many aspects of their ancient lifestyles as they can against the onslaught of modernity. For an excellent review of tribes of India see Fürer-Haimendorf (1982).

One might naively expect that since the sky is common to everyone and many of these Indian communities share a similar landmass and present-day lifestyle as settled farmers, their astronomical beliefs would be largely similar. However, in an earlier study (Vahia and Halkare, 2013) we showed how this assumption is wrong when we described the astronomical beliefs of the largest of the ancient Indian tribes, the Gonds. In the present study, we extend this survey to two more tribes of India—the Banjaras and the Kolams—who share the same landmass with the Gonds but have quite different histories when it comes to the amount of time that they have lived settled lives. The Gonds have been living in settled communities for centuries, while the Banjaras only began a settled existence about a hundred years ago and the Kolams only in recent memory. In Figures 1 and 2 we show the geographical distribution of the two communities.

In this paper we first discuss the histories of the two communities. Then we report on their astronomical systems before comparing these with each other and with Gond astronomy. However, there are some specific terms for astronomical constellations and asterisms that are unique to India, and which we will refer to. In particular, the seven most prominent stars in Ursa Major (sometimes referred to as the Big Dipper) that look like a spoon are called *Saptarshi* (*sapta* is seven, and *rishi* are saints).  In the interest of readability, we have listed our records of individual villages in the Appendices and only presented our main observations in the body of the paper.

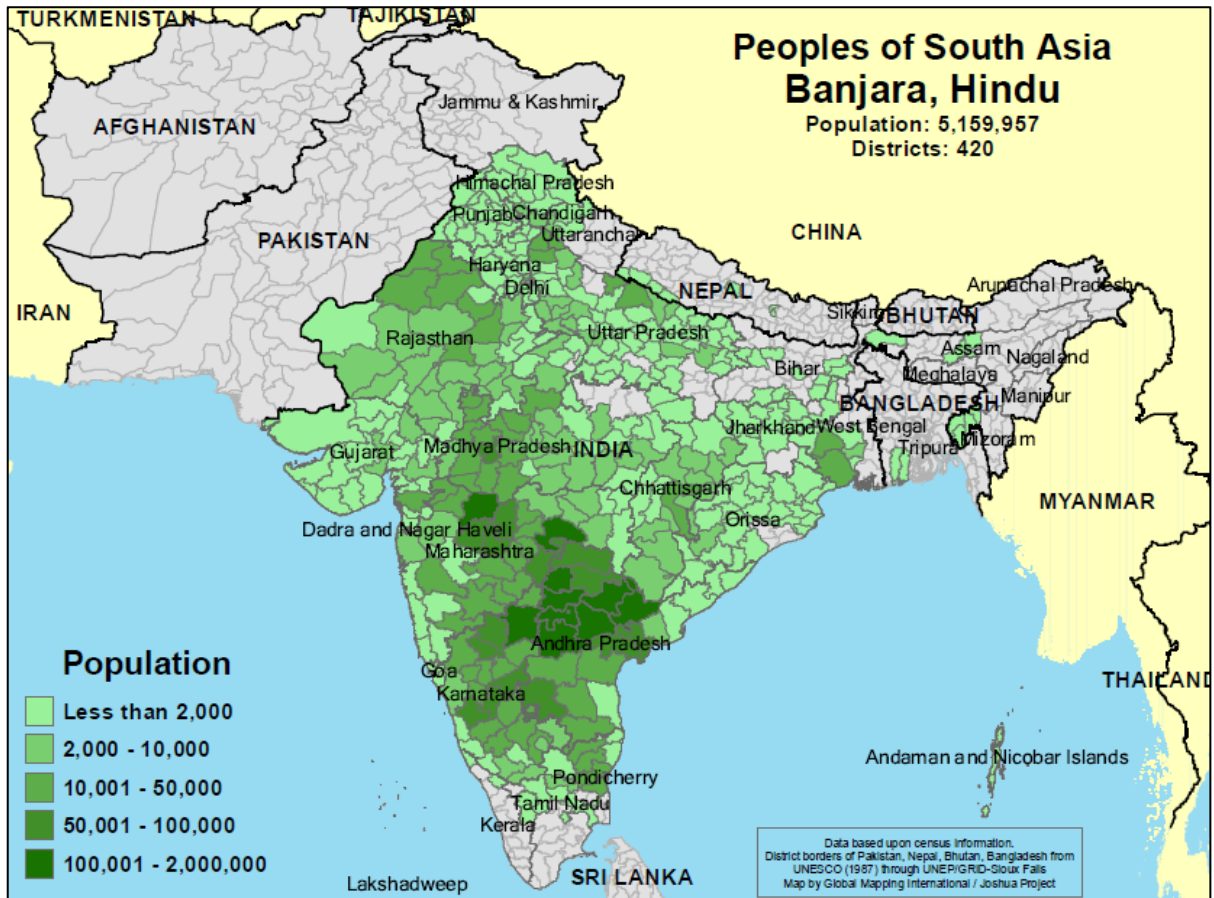

Figure 1: Geographical distribution of the Banjaras (courtesy: http://www.joshuaproject.net/).

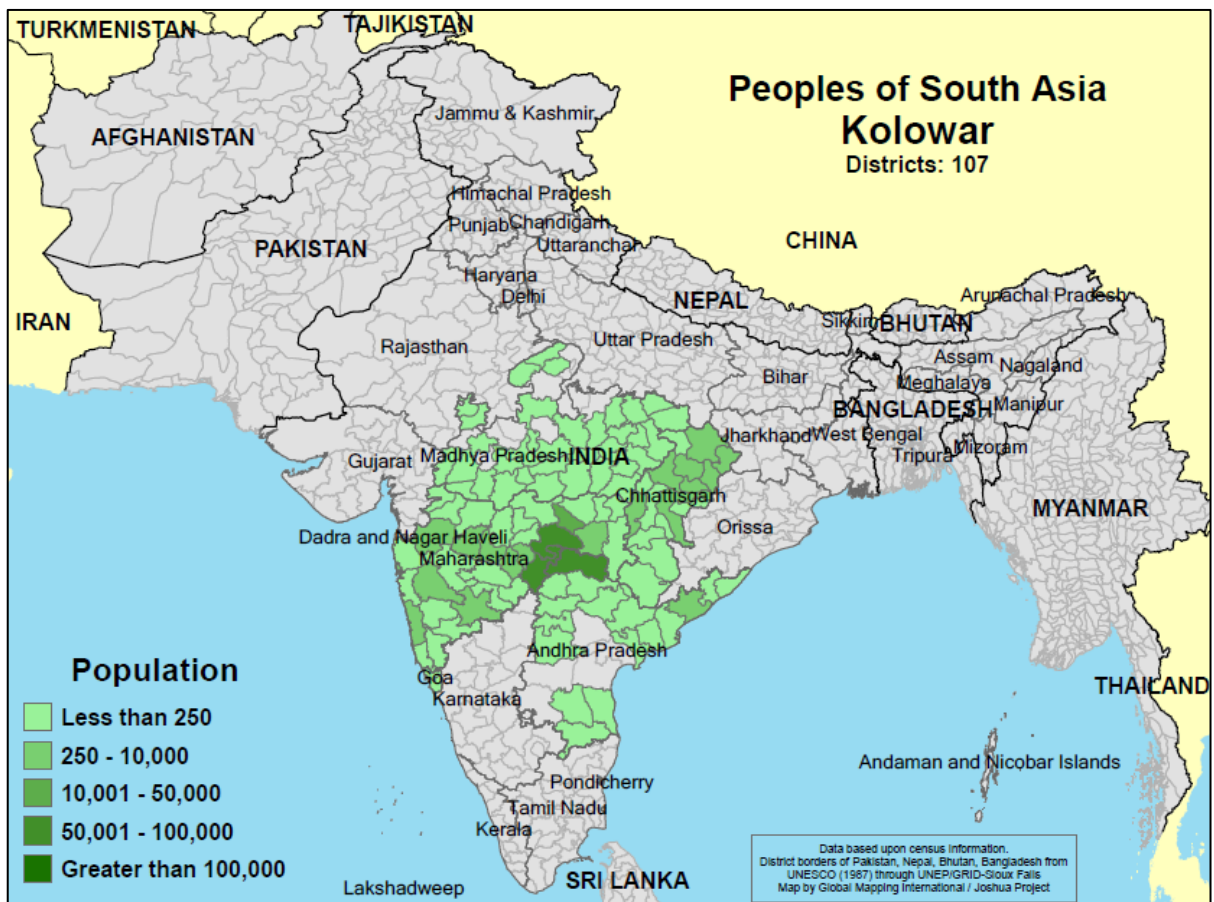

Figure 2: Geographical distribution of the Kolams (= Kolowars) (courtesy: http://www.joshuaproject.net/).

## 2 THE BANJARAS

The Banjaras are one of the largest tribal groups in India with a population of more than 5 million, and are found in large parts of the country (see Figure 1).  Traditionally they were connected with travelling caravans and the transportation of goods, and they traded over vast stretches of the Indian subcontinent.  They have a variety of names, such as Lambaras, Lambadas, which are subgroups within the larger Banjara population. The group we visited refer to themselves as Gor manus (people), and is one of the many subgroups of the Banjaras.  The word Banjara has several possible origins. The most likely of these are: from 'Ban', which means 'forest' (people who go into the forest), or from 'Baniya', which means 'trader' in the modern Hindi language (see e.g. Deogaonkar and Deogaonkar, 1992: 10).

There are two prevailing systems of beliefs amongst them: one group claims its roots from the (defeated) Rajputs of Rajasthan from the period of Prithviraj Chauhan (~1,200 AD), and the other group claims not to be Rajputs at all but instead to belong to a tribe whose roots go back to the Harappan period (~2,000 BC).  The latter are snake worshippers, and they say that this distinguishes them from the Sun- or Moon-worshipping Rajputs. Since snake-like ornaments are known amongst Harappan dolls, they use this connection to insist that originally they were traders from the Harappan period.  This is a potentially important link, since genetically they have an affinity to the gypsies of Europe (Mastana and Papiha, 1992) and the Harappans are known to have traded with West Asia.  There are also claims that the Banjaras are descendants of Luv, the elder son of the legendary Lord Rama of Ramayan.  For a summary of their lifestyle, see Deogaonkar and Deogaonkar (1992), and Vanjara et al. (2012). However, both groups agree that from ancient times, their primary profession was to move and trade goods throughout India.

In view of their profession, the Banjaras were not settled and were migrants across the vast Indian sub-continent, and until the arrival of the railways they were the sole transporters of goods across the country and beyond.  They moved in large groups of up to several hundred people, with herds of cows and bullocks loaded with goods.  They made very little use of horses, and traded bullocks for a very wide variety of goods, from gold to salt. They stayed only for a few days at any one place, except during the monsoon seasons when they preferred to stay at the same place for several weeks. Some Banjaras believe that their ancient currency was metal. One of their important items of trade was salt, which was obtained from coastal areas and then was traded in the hinterland.  This has left such a strong mark on them that even today, deep in the hinterland, some 500 km from the nearest sea coast, they still recall trading in salt.

The Banjaras only settled in permanent residences about one hundred years ago.  They call their villages '*Tanda*', a reminder that technically they are large moving groups that have set up a temporary residence. However, they do not seem to have used astronomy for navigation and used forward patrols and route markings to find their way about.

With their extensive experience in commerce, it is not surprising that politically the Banjaras are one of the most advanced of the tribes of India (Deogaonkar and Deogaonkar, 1992: 9). Over the last sixty years, at least one Chief Minister of the State of Maharashtra was a Banjara, and several Ministers in the State of Maharashtra, in neighbouring states, and in the national Governments of India have been Banjaras. They have also adapted to using the internet and other technology to maintain and propagate their identity, and they have a strong presence on the web (see e.g. www.banjaratimes.com/index2.html).

The Banjaras have seven primary goddesses, with their own mode of travel.  They are.

1) *Tulaja* or *Shitala* (the Goddess of Smallpox) (who uses a tiger)
2) *Maryama* (a lion)
3) *Mataral* (an elephant)
4) *Kankali* (a buffalo or bull)
5) *Hanglaj* (a horse)
6) *Chimar* or *Chimad* (a horse)
7) *Shyamki* (carried on a palanquin by humans).

Apart from these, they have two great saints: *Sevalal Maharaj* (the common greeting is "Jai Seva- Be Seva", meaning serve people, or Hail Sevalal Maharaj) and *Jetalal Maharaj*.  They largely follow only those Hindu festivals that are related to farming and their specific lifestyle, but even these have been modified so that now they have their own unique style.

The names of the month used by the Banjaras derive from modern Indian languages. Their main festivals are *Diwali*[1] and *Holi*. On *Diwali*, they worship the goddess of wealth—a very modern concept. However, the Banjaras celebrate it differently: on the morning of *Diwali* they have a feast for the ancestors. They also built up mounds of cow dung and plant wheat on the day after *Diwali*. In July, in a festival called *Pola*, cows are decorated and worshipped, a concept that has its roots in rural agrarian culture. *Holi* is a spring festival celebrated in late March, and it is followed by the marriage season, from *Chaita* (around April) to *Ashad* (June) —before the monsoon season. During the New Moon of *Akhadi* (July), a goat is sacrificed in anticipation of a good monsoon. When young girls come of age, there is the festival of *teej* in August/September when they sow wheat seeds (as a symbol of the goddess of fertility) in a bowl for eight days. The family then celebrates the ninth day when the goddess is merged with the soil. After this, the next festival begins on the eighth day of *Ashwin* (October) during the *Navratra* festival (a festival of nine nights, celebrated at the end of the farming season) when, again by local tradition, a goat is sacrificed. The last day of *Navaratra*, called *Dasera*, is in October (on the tenth day after the New Moon), and it is also celebrated with fanfare. Most of this information was given to us by the Bajaras we interviewed, but Deogaonkar and Deogaonkar (1992: 40-44) also discuss the festivals of the Banjara, though in a somewhat sketchy manner. It is probably no coincidence that the dates and times of Hindu festivals coincide with those of the Banjaras since the festivals of both communities arise essentially from farming-related rituals marking the sowing and harvesting seasons and other periods. The exact dates of the celebrations coincide with the Full or New Moon closest to the change of the seasons and may have been adjusted a little each year.

### 2.1 Genetic Data on the Banjaras

Genetic data on the Banjaras are sparse, possibly because their group identity is difficult to define. Studies such as the one carried out by Sachdev (2012) have focused on the Banjaras of the Rajasthan region, where they conventionally are believed to have originated. Mastana and Papiha (1992) have compared the genes of the Banjara sub-tribe called Lambana with the gypsies of Central and Western Europe. Their results suggest that gypsy populations of Eastern Europe still have great affinity with Indian nomadic groups, and the genetic differentiation in these populations may primarily be due to isolation, high rates of migration of subgroups towards Europe and genetic drift. The Western gypsies are more homogeneous as a local population, which may have resulted from a high degree of admixture.

In a recent study, Moorjani et al. (2013) compared the Roma population with various South Asian groups. They estimate that the Roma harbour about 80% West Eurasian ancestry— derived from a combination of European and South Asian sources—and that the date of the admixture of South Asian and European ancestry was about 850 years before the present. They provide evidence for Eastern Europe being a major source of European ancestry, and North-West India being a major source of the South Asian ancestry in the Roma. By computing allele sharing as a measure of linkage disequilibrium, they estimate that the migration of the Roma out of the Indian subcontinent was accompanied by a severe 'Founder Effect', which appears to have been followed by a major demographic expansion after their arrival in Europe. A Founder Event is when a very small group separates from a parent group carrying some vagaries in their genetic signal which then persist in the new community even though these provide no specific survival benefit. This may include facial features, body size, etc. (see e.g. Stone and Lurquin, 2010: 113).

### 2.2 Banjara Astronomy and Meteorology

The list of Banjara villages visited by us and their individual memory of astronomy and other information provided by them is given in Appendix 1, while the results are summarised in Table 1 below. Village 5 is not included in this Table since at that village we met a historian who supplied us with many details of the Banjaras but no insight into their astronomical knowledge and beliefs.

The most commonly-known astronomical objects amongst the Banjaras are the stars in Orion (which they see as a deer), and the Pleiades asterism, which they proudly proclaim to be a piece of jewellery that is worn on the forehead and that typically has many little metallic balls (generally of silver) strung together to appear like a bunch of grapes. They know of an evening star and morning star—no particular star, just the one that tells them that the day or the night has just begun. They have many indicators to predict the intensity of the monsoon, the two most favoured being the glow around the Moon, and the activities of the crow: if it builds its nest high up in a tree in late May strong rains are unlikely but if it builds a well-protected nest in the lower branches then the rains are expected to be heavy. The direction from which the rains will come is opposite to the direction in which the nest is made in relation to the tree trunk.

The Banjaras are aware of comets as stars with tails, and they think of them as bad omens. Meteors also are considered bad omens. *Saptarshi* (the Big Dipper) is divided into the four stars of the polygon that form the death bed (*Jamkhat*), with a procession of three people following the bed. Upon dying, Banjaras walk along the Milky Way (the path of the dead) to reach the heavens. They do not seem to have stories about the afterlife, and the worship of ancestors is more in the sense of gratitude than out of any expectation (unlike the belief in main-stream Hinduism).

Table 1: The astronomical knowledge and beliefs of the Banjaras.

|  | Knowledge or Belief | Banjara Village ||||||
|---|---|---|---|---|---|---|---|
|  |  | 1 | 2 | 3 | 4 | 6 | 7 |
| 1 | The Sun: *Dan* or *Dado* | x |  |  |  |  | x |
| 2 | The Moon: *Chanda* – similar to the Sanskrit name | x | x | x | x |  | x |
| 3 | The lunar maria: a tree or an old lady weaving cotton under a banyan tree | x |  | x |  |  | x |
| 4 | Eclipses are recognised | x |  |  |  |  | x |
| 5 | A meteor: *Tara tutgo* (a broken or falling star; usually a bad omen) | x | x | x | x |  | x |
| 6 | A comet: *Seshar Tara* (a smoking star or star with a tail; a good or bad omen) | x | x | x | x |  | x |
| 7 | The Milky Way: *Mardaar wat* (the path of the dead or the path of animals) | x | x | x | x |  | x |
| 8 | The Bigger Dipper: *Jamakhat* or *Yamakhat* (the cot of the dead) | x | x | x | x | x | x |
| 9 | Canis Major: *Medi* – connected with processing the harvest | x |  |  |  |  |  |
| 10 | The Pleiades: *Shirser Jhumko* or *Jhumko tara* (jewelry worn on the forehead) | x |  |  | x | x | x |
| 11 | Orion: *Halni*, *Halini* or *Harini* (a deer) | x | x | x | x | x | x |
| 12 | Taurus: *Kamedi* (a dove), close to *Halni* |  |  |  | x |  |  |
| 13 | The morning star: *Porya Tara* | x | x | x | x | x |  |
| 14 | The evening star: *Subtara* (a good omen) |  |  | x | x | x |  |
| 15 | The Pole Star is identified | x | x | x |  |  |  |
| 16 | A gnomon was used to determine time |  |  |  |  | x |  |
| 17 | An intercalary month (*Dhonda Mahina*) synchronises solar and lunar calendars | x |  |  |  |  |  |
| 18 | Names exist for the four cardinal directions |  | x |  |  |  |  |
| 19 | The monsoon is predicted by the size of the glow around the Moon | x | x | x | x |  |  |
| 20 | The monsoon is predicted by the direction of lightning |  | x | x | x |  |  |
| 21 | The monsoon is predicted by the location of crows' nests |  | x |  | x |  | x |
| 22 | The monsoon is predicted by the sighting of Taurus in the east | x |  |  | x |  |  |
| 23 | The monsoon is predicted by the direction of clouds |  | x |  |  |  |  |
| 24 | A bad monsoon is predicted if ants are coming out of the ground with their eggs |  |  |  | x |  |  |
| 25 | A bad monsoon is predicted if it hails |  |  |  | x |  |  |
| 26 | A rainbow indicates the end of the current rains |  |  |  | x |  |  |

## 3 THE KOLAMS

The Kolams (also known as the Kolowars) are a relatively small tribal group. They are found largely in south central India (Figure 2), and their total population at around 400,000 is less than one tenth that of the Banjaras. Until as recently as the 1940s and 1950s they typically practiced slash and burn farming, and they would beat the earth by hand using a wooden stick to soften it rather than use a plough. They largely relied on foraging for their survival. For an extensive review of the Kolams see Deogaonkar and Baxi (2003). They were, therefore, renowned for their familiarity with the jungle and skill in divination and the propitiation of local gods. This reputation has led many Gond communities to entrust the cult of certain local divinities, and particularly those of the gods holding sway over forests and hills, to the priests of nearby Kolam settlements. It is because of this sacerdotal function of the Kolams that the Gonds refer to this entire tribe as 'Pujaris' or priests (Fürer-Haimendorf, 1982: 13). The Kolams refer to their habitations as *Pods*. Fürer-Haimendorf (ibid.) further notes that till the 1940s Kolams (and a closely-related group, the Naikpods) practised slash-and-burn cultivation using hoes and digging sticks. Currently, under the policy of the Government of India to preserve forest areas and clear them of human habitation, only small numbers of Kolams live in hill settlements. Now, most of them are found in villages on the plains, where they work as tenant farmers or agricultural labourers. Some of them own the land they cultivate. They are scattered over a large area. Kolams and Naikpods originally had languages which closely resembled each other but have been diluted by the onslaught of modernity.

Where Gonds and Kolams lived in close proximity, the Gonds usually settled at the foot of the higher ridges and cultivated the valleys, plateaux, and gentle slopes, while the Kolams built their hamlets on ridge tops and cultivated the steep hillsides. The crops sown and reaped consisted mainly of small millets, sorghum, maize, and certain vegetables such as beans, taro, and marrows. This provided a family with sustenance only for about seven to eight months the year, and during the remaining months wild fruits, herbs and roots formed the mainstay of their diet (Fürer-Haimendorf, 1982: 85).

The Kolams have five primary Gods (*Ayak*): *Bhimayak, Jangobai, Nadadi Amma* (village goddess), *Vaanadeya* (forest god) and *Sandun* (Sun) who is also called *Suryak* (Sun).

The Kolams divide the year into twelve months (conventional months are lunar with an intercalary month every three years). The corresponding modern calendar months, listed below, are approximate since the Kolam calendar is lunar in nature, starting with Full Moon, as used by the Kolams.

- *Bhavai – Devkaran* – May
- *Bud bhavai Nela* – June

- *Aakhadi Nela* – July
- *Pora* – August
- *Petala* – September
- *Divala* – October
- *Konka divala* – November
- *Sati Nela* – December
- *Pusi Nela* – January
- *Bhimrasi* – February
- *Duradi* – March
- *Saita* – April

Their counting system is decimal, with names similar to the Dravidian numerals with numbers 1 to 10 given as *Okkad, Indi, Mundi, Nale, Eda, Aar, Ed, Enmdi, Tomdi* and *Padi*. They have no special name for zero.

### 3.1 Genetic Data on the Kolams

Rao et al. (1992) have reported a study of several Gond tribes and Kolams from Maharashtra, along with other data such as the level of endogamy etc. On the basis of these data, they show that Kolams have historically lived in isolated groups with a high level of endogamy (31%) but have genetic similarity to Raj Gonds (see also, Majumder and Mukherjee, 1993). This is expected due to the fact that they have historically served as priests to the Gonds (Fürer-Haimendorf, 1982: 13). However, Sachdeva et al. (2004) reporting on the Kolams of Andhra Pradesh show that the genetic distance between the two is relatively large. This seems to suggest that the contact between the two tribes is sensitive to the region of interaction. After studying the Kolams of Andhra Pradesh, Sirajuddin et al. (1994) concluded that they seem to have remained isolated even to the present day with very little new gene inflow into the population.

### 3.2 Kolam Astronomy

The list of Kolam villages visited by us is given in Appendix 1, along with their astronomical and other knowledge. Salient features of their astronomical beliefs are given in Table 2

The most commonly-known asterism among the Kolams is the Pleiades which they see as a collection of one large and several small birds. To them, *Saptarshi* is a cot which three stars (a Kolam, a Gond and a *Pradhan* – chieftain) are trying to steal.

They also have a fairly detailed myth about solar eclipses. A total solar eclipse is considered to be a good omen. A partial solar eclipse where the part of the Sun towards the zenith is covered, is a bad omen for humans, but if the part of the Sun towards the horizon is covered it is inauspicious for animals. The underlying idea is that the tax-collecting gods have come to collect their due and if they can get their complete share (totality) they are satisfied; otherwise they make up the deficit by making animals or humans pay. Two villages also discussed lunar eclipses as having different shades, the lighter shade (red) arises when Moon is being run over by a caterpillar while the darker shade (black) suggests that a frog is eating up the Moon.

Another detailed and commonly-held story is that of an asterism called *Samdur*, which we identify as the Great Square of Pegasus. The reason for this is that apart from the direction pointed out by them, *Samdur* means both the sea (badly pronounced) and an object where all points are equidistant. Pegasus rises in the east soon after midnight in June. While showing the asterism the Kolams also consistently pointed to the east. We therefore searched for an asterism that could meet this description. Pegasus fits this description well. It is a great lake that provides water for the land. This constellation was studied in detail, and five different animals come to drink and feed there. They are a frog, buffalo, deer, horse and peacock (and, on occasions, a pig). These animals all predict up-coming rain: the peacock and deer produce average rain while all of the other animals suggest a good monsoon. Of the other animals, an especially clear sighting of the frog is very auspicious. We show this in Figure 3, where we have positioned each animal based on our understanding of the patterns that best fit that specific animal. As it turns out, the two animals that are further east of Pegasus indicate a bad monsoon and the asterisms that are to the northeast of Pegasus indicate a good monsoon. An alternative explanation is that *Samdur* refers to Centaurus, which is seen in the south in June. In this case, the fit for the different animals is shown in Figure 4.

Another unique story is that of Cygnus where they identify three pots placed on top of each other. They know of Scorpius as a snake and probably include some stars above the conventional Scorpius in this.

In Crux—the only southern constellation they recognise—they can see the *Mahua* tree just as the Gonds do. They see Taurus as a bird with two eggs. They know only the inner stars of Orion as the belt of a hunter, and the region around the Orion Nebula as a sword which may be a weak memory of the original Hindu myth which sees all of Orion as a deer and the three stars of the belt as an arrow. On the Moon, they can see an old lady sitting under a tree and spinning cotton.

An interesting aspect of the 270 solar eclipses that the Kolams could have seen between AD 1000 and 2000 is that only one was annular (on 30 December 1758) and one was total (on 22 January 1898) as per the Eclipse Predictions of Fred Espenak and Chris O'Byrne (NASA's GSFC: (http://eclipse.gsfc.nasa.gov/eclipse.html). It is probably because of this that the Kolams place so much emphasis on partial eclipses.

Table 2: The astronomical knowledge and beliefs of the Kolams.[2]

| | Knowledge and Belief | Kolam Village | | | | | | | | | |
|---|---|---|---|---|---|---|---|---|---|---|---|
| | | 1 | 2 | 3 | 4 | 5 | 6 | 7 | 8 | 9 | 10 |
| 1 | Solar eclipses: They consider these are payments of dues by the Sun – a partial eclipse means partial payment and the rest has to be paid by either humans (if the eclipsed section is in the top half of the Sun) or animals (if the eclipses section is in the bottom half of the Sun) | | x | x | x | x | x | x | x | x | x |
| 2 | Lunar eclipses: light red colour if a caterpillar or scorpion is eating the Moon and dark red if a frog is eating it | | | | | | x | | x | x | x |
| 3 | A comet: *Sipur suka* (a star with a tail; a bad omen) | | x | x | | | x | | x | x | x |
| 4 | A meteor: *Suka erengten* (a broken star, falling star or a star with a tail; stellar excreta; a bad omen) | | x | x | x | | | | | | x |
| 5 | The Milky Way: *Margam*, *Panadi*, *Bhori Sanko* or *Jev* (an animal path) | x | x | x | x | x | | | | x | x |
| 6 | The Big Dipper (four stars in a quadrangle): *Mandater* (a cot) | x | x | x | x | x | x | x | | x | x |
| 7 | The Big Dipper (the three trailing stars): a Kolam, a Gond and Pardhan (chief), starting from the cot | x | x | x | x | x | x | x | | x | x |
| 8 | Crux: *Irukmara* or *Ipamaka* (a *Mahua* tree whose flowers are fermented to produce alcohol); α Centauri is *Murta* (an old lady) and β Centauri is *Kadma* (a young lady), both there to collect flowers. | | x | x | x | x | | x | x | x | |
| 9 | Cygnus: *Kavadi Kunde* or *Kavedi Koda* (a tower of three pots) | x | | x | x | x | | x | x | | |
| 10 | *Danedare Pila*, a spiral-shaped asterism: probably the tail of Scorpius | | x | | | | | | | x | |
| 11 | Great Square of Pegasus: *Samdur* (a sea from which rain water comes) | | x | | x | x | x | x | x | x | x |
| 12 | Stars around Pegasus: the following animal asterisms are seen around Pegasus: a peacock, buffalo, frog, deer and horse, and sometimes also a pig; their relative brightness decides the intensity of up-coming rain | | x | | | x | x | x | x | x | x |
| 13 | Leo: *Murda* (a dead body – this is a Gond myth) | | x | | | | | | | x | |
| 14 | Orion: *Tipan* or *Trivpate* (an instrument used to sow seeds) | x | x | x | x | x | x | x | | x | x |
| 15 | The Pleiades: *Kovela Kor* (one large and several small birds) | x | x | x | x | x | x | x | x | x | x |
| 16 | The Pleiades: *Sappa* (an instrument used to process the harvest) | x | | | | | | | | | |
| 17 | Scorpius: *Borenagu* or Nago (a snake), or *Nagun* (a cobra) | | | x | x | x | x | x | x | x | |
| 18 | Scorpius: *Tuntor* or *Tootera* (a scorpion) | x | x | | | | | | | x | x |
| 17 | Taurus*: Bhori*, a bird (Aldebaran) with two eggs | x | | x | | | x | x | x | | |
| 19 | The morning star: *Vegud suka* (star of the morning) | x | x | x | x | x | x | x | | x | x |
| 20 | The evening star: *Jevan suka* (a star of dinner time) | x | x | x | x | x | x | x | | x | x |
| 21 | Sirius: *Met* (a pole to which a bull is tied and moves in a circle on the ground (*Kalave*) to separate the rice from the husk) | x | x | x | | x | | | | | |
| 22 | Names exist for the four cardinal directions | x | | | | | | | | | x |
| 23 | The monsoon is predicted by the size of the glow around the Moon | | | | | x | | | | | x |
| 24 | A rainbow: *Ayak* or *Aikawa* (indicates the end of the current rains) | | | | x | | x | | x | x | |

## 4 CONCLUDING REMARKS: A COMPARATIVE STUDY

Here we provide a comparative discussion of the astronomical beliefs of three different Indian tribes in relation to their evolving survival needs and inter-tribal influences. The three tribes in question are the Gonds (see Vahia and Halkare, 2013), along with the Banjaras and the Kolams, whose astronomy is reported here. It should be kept in mind that we have studied the three tribes in regions where they are spatially co-located, but all three tribes in general have a much wider geographical distribution. Table 3 provides an overview of the three communities.

In general, most of the astronomical observations made by these communities fall into two groups. The first set refers to the period just before the monsoon and the asterisms noted are connected with the expectation of monsoonal rain (March to May), while the other set of observations were made immediately after the monsoon (i.e. September to December).

Amongst the three tribes, the Gonds are the most settled and probably the most ancient tribe, with a continuing history, farming traditions and settlement dating back three thousand years or more. Against this, the Banjaras are more recent settlers. While they trace their origin to *Prithviraj Chauhan* (12[th] century AD), they are probably a much older group whose profession was that of migrants and traders (see e.g. Deogaonkar and Deogaonkar, 1992: 12). They share common traits with the gypsies of Europe. Some of them insist that the Banjaras date back to the Harappan period when they were the traders who took goods from the Harappan Civilisation all the way to West Asia and beyond, as well as within the Indian subcontinent. They were forced to settle down when their main profession as transporters and traders was threatened with the arrival of railways in India more than one hundred and sixty years ago. However, having been hardened by millennia of travel, they proved to be a robust community which quickly and aggressively took to settled life across the sub-continent. Since then, Banjaras have held many important administrative and political posts in India. The Kolams, the smallest of the traditional tribes considered here, essentially survived as foragers with only the simplest form of farming. It is only in the last few decades that the Indian Government's aggressive attempts to settle them have made them adopt various trappings of citizens of a modern state. In the course of our research, we met some elders who had been foragers as youngsters and have now settled down on the land given to them by the Government.

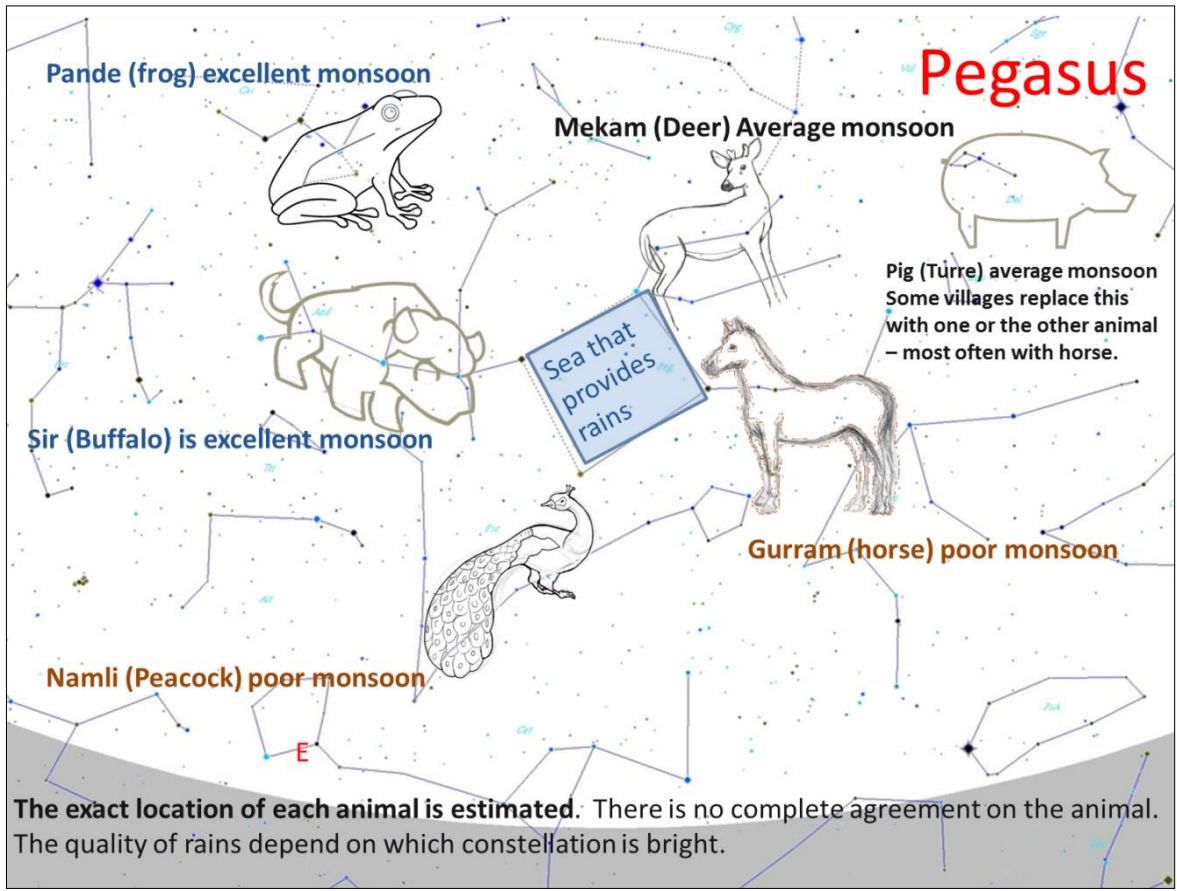

Figure 3: *Samdur* as Pegasus, the great sea that provides the rains, and the five or six associated animals.

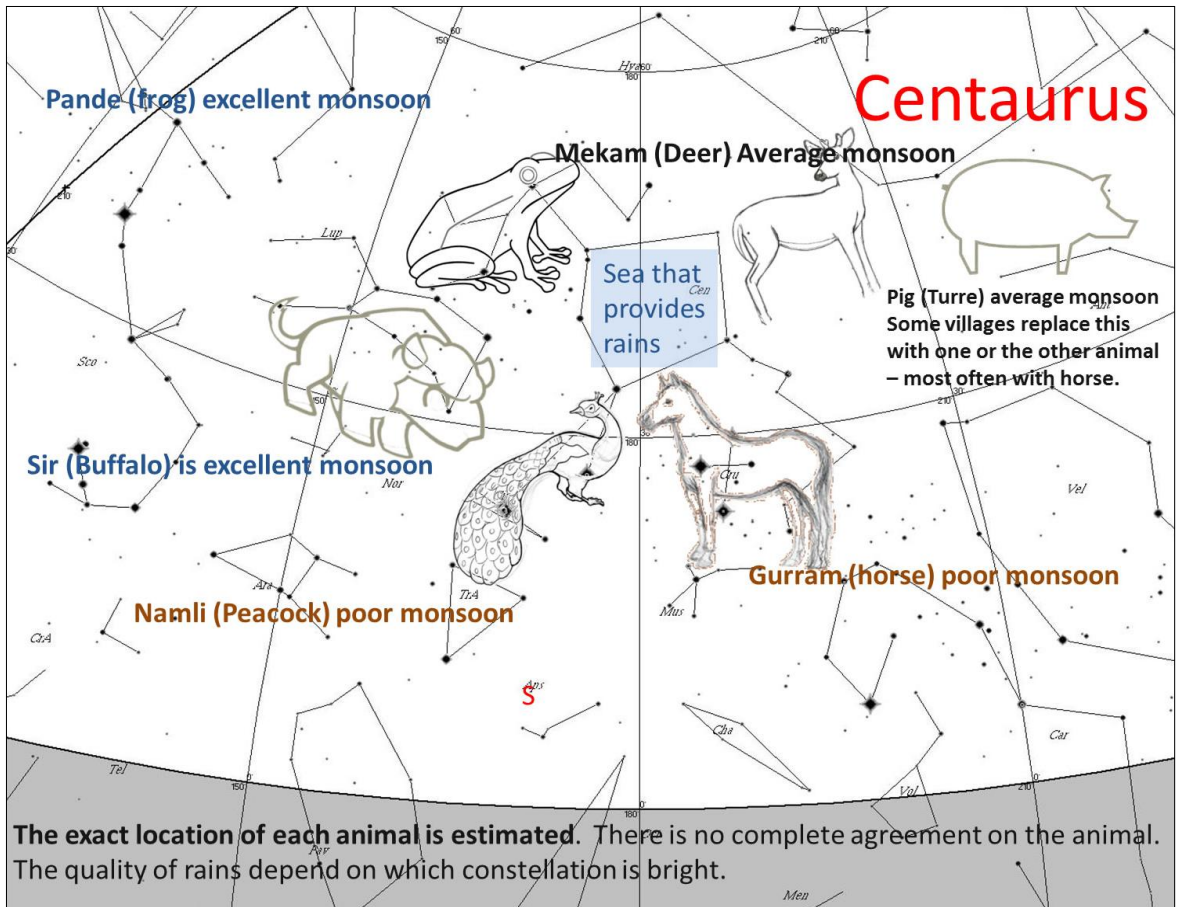

Figure 4: The alternative explanation, that *Samdur* may represent Centaurus,

An interesting feature of these tribes is their strong sense of identity, aggressive desire to live in isolation from other nearby communities, and as far as possible not interact in any way. This has allowed the three communities to retain their traditional genes, languages and cultural traditions even though modern education has produced severe stress on all of these aspects. As we shall see below, their strong sense of tribal identity has allowed each group to create its own astronomical myths and stories. The complexity of their astronomical ideas and beliefs is directly related to the period during which they have lived a settled lifestyle.

Their astronomical knowledge, especially the number of markers they use to predict the monsoon, suggests that native astronomy is a continuing tradition. They continue to rely on their beliefs and predictions that are continuously evaluated and modified (or more markers are added) to improve their predictions. Another pattern that emerges is that just like seasons, their association of the sky with philosophical ideas is also related to the period that they have spent in a settled existence. These particular wanderers, it would seem, have little interest in philosophy and astronomy. For example, even though the Banjaras were travellers and movers of goods, they did not use astronomy for travel, and relied entirely on local knowledge and markers left behind to guide them.

Table 3: A comparison of the social aspect of the three Indian tribes studied by us.

| Property | Banjara | Kolam | Gond |
|---|---|---|---|
| Population size | A large ethnic group spread over all of India. Population more than 5 million. | A very small primitive tribe. Population less than 0.5 million. | The largest ethnic group. Population about 10 million. |
| Geographical spread | All India | Central India | Central India |
| Language spoken | Austro Asiatic – a mix of Rajasthani and Gujarati with a sprinkling of local adaptation (Deogaonkar and Deogaonkar, 1992: 57). Language code: ISO 639-3 lmn | Dravidian, with some primary forms of Telugu and Kannada language (Deogaonkar and Baxi, 2003: 38). Language code: ISO 639-3 kfb | South Central Dravidian. Language code: ISO 639-2 gon ISO 639-3 gon – inclusive code Individual codes: ggo – Southern Gondi gno – Northern Gondi (From Wikipedia) |
| Original profession | Trading and transportation | Foraging, hunting | Farming |
| Current profession | Farming | Farming | Farming |
| Gods | One primary goddess + seven other goddesses. | One primary goddess, four *Deva* + natural forces. *Maruti* temples were also seen. There are separate village *devis* (goddesses). | Nine gods + *Lingo*, *Jingo*. Temples contain iron tools and weapons. |
| Approach to nature | Indifferent; they rely on terrestrial signals to predict the upcoming monsoon. | Sensitive to all aspects of nature, particularly in foraging. | Very reverential and sensitive. |
| Astronomy | Simplest astronomy that barely goes beyond the beliefs of others except for finding jewelry in the sky. Surprisingly, they do not use astronomy for travel. | Extensive, if a somewhat generalised visualisation of the sky in the sense that they do not seem to record changes in the sky over the year, and they recall the sky only during the immediate pre-monsoon period when it is well observed. | Fairly extensive. Regarding farming, they have a myth that Ursa Major was not allowed to sleep, suggesting that they remember it being circumpolar. From precession calculations it can be shown that this happened more than 3,000 years ago. Stories of all aspects of life are found in the sky. |
| Predictive astronomy | Not very sensitive. | Pegasus delivers water. | Extensive and used even today. |
| Treatment of dead | Buried the dead and then they were forgotten. Ancestors were worshipped as a group twice a year. | Buried the dead but remembered them regularly every year and on important occasions. Graves marked by stones. | Buried and remembered as a group. Graves marked by stones. |
| Stories of origin | Firmly remember a relation to the Rajputs. Others claim a an Harappan ancestry. | Cannot recall much. Worked as priests for the Gonds in AD 1200. | Stories go back to 1000 BC. |
| Festivals | Farming and related festivals. | Farming and related festivals. | Farming and related festivals. |

In Table 4 below, we list the observations reported in more than just one village. The list is in decreasing order of frequency of report, but we ignore minor local fluctuations which are faithfully reported in the Appendices. The principle differences and similarities that appear are as follows:

1) Orion and Taurus are seen collectively as a farming scene by the Gonds while imagination is more nebulous amongst the other tribes. To the Gonds, Orion is the plough, Taurus is a bird trying to eat the seeds, and the Pleiades is a bunch of stones thrown to chase it away. The eastern end of Orion and Canis Major complete this scenario, with other farming-related activities. The Kolams recognise the three stars in the Belt of Orion as farming equipment, and see Taurus as a bird with two eggs. In addition, they vaguely recall that in the Big Dipper there is a facility with other farming activity. In marked contrast, the Banjaras barely notice Orion and Taurus.
2) The Pleiades asterism is seen as a piece of jewellery by the Banjaras, a group of birds by the Kolams and a bunch of stones thrown at Taurus by the Gonds.
3) All three tribes rely on the glow around the Moon prior to the monsoon season to predict the intensity of the upcoming rains.
4) A comet is a star with a tail for the Banjaras, while the Gonds see it as a broom of the Gods. The Kolams also mention the broom and tail but are not very specific about its effect upon humans.
5) The Kolams have a deep-seated vision of the great lake in Pegasus and the set of animals coming there to drink, which they use to predict the monsoon. The two other tribes ignore Pegasus, although the same Kolam word '*Samdur*' is used by the Gonds to identify Auriga.
6) Similarly, to the Kolams Cygnus is a set of three pots stacked one on top of the other —an idea that is completely alien to the other two tribes.
7) Only the Banjaras could identify the Pole Star.
8) The Banjaras ignore Scorpius, while the Kolams and the Gonds see this constellation as a snake or a scorpion.
9) The most interesting differences occur in the perception of eclipses. Both the Banjaras and the Gonds ignore eclipses while the Kolams have elaborate stories to explain solar and lunar eclipses.
10) The Banjaras use the crow's nesting habits at the end of May to predict the onset of the monsoon.
11) Both the Gonds and the Kolams know of Crux as the *Mahua* tree, while the Banjaras ignore it.
12) The Banjaras think that the Milky Way is the path dead humans follow, while for the Kolams and the Gonds it is an animal path.
13) In the case of S*aptarshi* (the Big Dipper), the Banjaras consider it to be a cot of the dead while the Kolams have a vague idea of some attempts to steal it. The Gonds are very clear that the legs of the cot are made of precious metals and therefore are worth stealing.
14) One interesting, but surprisingly rare story supplied by just one Kolam village is that Ursa Major is a container with water, which is emptied during the monsoon. The failure of this imagery to gain popular acceptance may be because Ursa Major appears as a spoon facing downwards for a much shorter period than the length of the monsoon season. The monsoon comes to this region in mid to late June. The front two stars of *Saptarshi* (the most prominent stars of Ursa Major) are aligned to Orion. Hence when Orion rises, the container is vertical to the horizon and when Orion is in the western horizon, the container seems to empty itself, a scene that would be witnessed in the sky at sunset in June.
15) Leo as a procession of death is only known to the Gonds. Meanwhile, Crux, Cygnus and Pegasus are only known to the Kolams.
16) All three tribes have the idea that dinner should start when the first star is seen and that the day should begin when the morning star is seen.

In Table 5 we list the differences between the three tribes for the same astronomical entity. As can be seen from the table, even spatially-overlapping communities tend to have dramatically-differing knowledge of and interest in astronomy.

Most of the observations of the sky were made from March till July, close to the monsoon season. These tribes seemed to assume that the sky was the same at other times of the year or they do not bother to look at the sky except in the post-summer period when they needed to start worrying about the rain and sowing their crops. This indifference towards the sky is also interesting in the sense that even though it must have been visually striking, these tribes did not seem to have been impressed by it as a matter of curiosity. For example, they do not seem to have kept track of the movement of the rising (or setting) point of the Sun on the horizon in the course of the year, which contrasts markedly with southern India where megalithic solar observatories abound (see Menon et al., 2012).

Also, traditionally the Kolams were foragers, and they seem to have a lot more to say about the sky in a nebulous manner. The long tradition of farming amongst the Gonds is equally evident in their beliefs of all aspects of astronomy and religion. The general indifference of the Banjaras to the sky (except for the Pleiades being seen as a rich piece of jewellery) is also apparent as a result of our survey. Clearly, the relationship of man to the sky is a reflection of his relationship to the land.

Table 4: Comparison of astronomical and meteorological perspectives of the tribe tribes. The entries in red print indicate the unique features of that tribe. The entries are listed by the frequency of their appearance in the different villages. Only information from two or more villages is listed here.s

| Banjara | Kolam | Gonds |
|---|---|---|
| Orion: *Halni*, *Halini* or *Harini* (a deer) | The Pleiades: *Kovela Kor* (one large and several small birds) | Orion: *Naagardai* (a plough); the belt of Orion: *Tipan* |
| The Big Dipper: *Jamakhat* (the cot of the dead) | The Big Dipper (four stars in a quadrangle): *Mandater* (a cot) | The Big Dipper (four stars in a quadrangle): *Sedona* (old lady's) *Katul* (cot) |
| The morning star: *Porya Tara* | The Big Dipper (the three trailing stars): a Kolam, a Gond and a Pardhan (chief), starting from the cot | The Big Dipper (the three trailing stars): three thieves trying to steal the cot, the legs of which were made of gold, silver and copper |
| The evening star: *Subtara* (a good omen) | Orion: *Tipan* or *Trivpate* (an instrument used to sow seeds) | The morning star: *Pahat Sukum* |
| A comet: *Seshar Tara* (a smoking star or star with a tail; can be either a good or a bad omen) | The morning star: *Vegud suka* | The evening star: *Jevan Sukum* |
| A meteor: *Tara tutgo* (a broken or falling star; usually a bad omen) | The evening star: *Jevan suka* | Comets: *Jhadani*, *Bhimal Saat*, *Kayshar*, *Jhadu*, *Bahari* (a weapon of the gods; a good omen) |
| The Moon: *Chanda* – similar to the Sanskrit name | Solar eclipses: They consider these are payments of dues by the Sun – a partial eclipse means partial payment and the rest has to be paid either by humans or by animals | The Milky Way: *Dhor Sari*, *Rasta*, *Sagur*, *Murana Marg*, *Pandhan*, *Sadak* (the great path of animal migration) |
| The Milky Way: *Mardaar wat* (the path of the dead or the path of animals) | Great Square of Pegasus: *Samdur* (a sea from which rain water comes) | The Sun: *Lingo*, *Purbaal*, *Bera*, *Vera*, *Din*, *Suryal* |
| The monsoon is predicted by the size of the glow around the Moon | Crux: *Irukmara* or *Ipamaka* (is a Mahua tree) | The Moon: *Jango*, *Chandal, Nalend* or *Nalen* |
| Pleiades: *Shirser Jhumko or Jhumko tara* (jewelry worn on the forehead) | Stars around Pegasus: the following animal asterisms are seen: a peacock, buffalo, frog, deer and horse, and sometimes also a pig; their relative brightness decides the intensity of up-coming rain during the monsoon season | The Pleiades: *Mogari*, *Mongari*, *Kutpari*, *Thengari*, *Mundari* (stones thrown at birds) |
| The Pole Star is identified | The Milky Way: *Margam*, *Panadi*, *Bhori Sanko* or *Jev* (an animal path) | Meteors: *Ulka*, *Sukir Pelkta*, *Tara Urungta* (generally these are considered to be stellar excreta, or occasionally souls that are falling from their holy places in the sky) |
| The monsoon is predicted by the direction of lightning | Scorpius: *Borenagu* or Nago (a snake), or *Nagun* (a cobra) | Canis Major: *Purad* or *Hola* |
| The monsoon is predicted by the location of crows' nests | Cygnus: *Kavadi Kunde* or *Kavedi Koda* (a collection of three pots stacked one on top of the other) | The monsoon is predicted by the size of the glow around the Moon |
| The lunar maria: a tree, or an old lady weaving cotton under a banyan tree | A comet: *Sipur suka* (a star with a tail; a bad omen) | Auriga: *Samdar* (this indicates the nature of the up-coming monsoon) |
| The monsoon is predicted by the sighting of Taurus in the east | Taurus: *Bhori* (a bird, Aldebaran, with two eggs) | Scorpius: *Michu* (responsible for producing the dead body, *Murda*, which is seen in Leo) |
| The Sun: *Dan* or *Dado* | Scorpius: *Tuntor* or *Tootera* (a scorpion) | Sirius region: *Topli* (the basket of seeds used when sowing the fields) |
| Eclipses are recognised | Sirius: *Met* (a pole to which a bull is tied and moves in a circle on the ground (*Kalave*) separating the rice from the husk) | Taurus: *Medi* (part of a large farming scene also involving Canis Major, Lepus and Orion) |
| | A meteor: *Suka erengten* (a broken star, falling star or a star with a tail; stellar excreta; a bad omen) | |
| | A rainbow: Ayak or Aikawa (indicates the end of the current rains) | |
| | Lunar eclipses: light red colour if a caterpillar or scorpion is eating the Moon and dark red if a frog is eating it | |
| | *Danedare Pila*, a spiral-shaped asterism: probably the tail of Scorpius | |
| | Names exist for the four cardinal directions | |
| | Leo: *Murda* (a dead body –a Gond myth) | |
| | The monsoon is predicted by the size of the glow around the Moon | |

Table 5: Differences in the perspectives of the three different tribes for the same astronomical object.

| Object | BANJARA | Kolam | Gonds |
|---|---|---|---|
| Orion | *Halni* | *Tivpate* | *Tipan* |
| Ursa Major | *Saptarshi*: *Jamakhat* (the cot of death) | *Saptarshi*: *Mandater* (cot) | *Saptarshi*: old lady's cot |
| Morning Star | *Porya Tara* | *Vegud suka* | *Pahat Sukum* |
| Evening Star | *Subtara* (star of good omen) | *Jevan suka* | *Jevan Sukum* |
| Ursa Minor | | The three following stars are three people, a Kolam, a Gond and Pardhan (chief) | The three following stars are three thieves |
| Comets | *Seshar Tara* (a smoking star or star with a tail; a good or bad omen) | *Sipur suka* (a star with a tail; a bad omen) | *Jhadani*, *Bhimal Saat*, *Kayshar*, *Jhadu*, *Bahari* (a weapon of the gods; a good omen) |
| Pleiades | *Shirser Jhumko or Jhumko tara* (jewelry worn on the forehead) | *Kovela Kor* (one large and several small birds) | *Mogari*, *Kutpari*, *Thengari*, *Mundari* (stones thrown at birds) |
| The Milky Way | *Mardaar wat* (the path of the dead or the path of animals) | *Margam*, *Panadi*, *Bhori Sanko* or *Jev* (an animal path) | *Dhor Sari*, *Rasta*, *Sagur*, *Murana Marg*, *Pandhan*, *Sadak* (the great path of animal migration) |
| Meteors | *Tara tutgo* (broken or falling stars; usually a bad omen) | *Suka erengten* (a broken star, falling star or a star with a tail; stellar excreta; a bad omen) | *Ulka*, *Sukir Pelkta*, *Tara Urungta* (generally considered to be stellar excreta) |
| Taurus region | *Kamedi* (the dove; if seen in the east will bring rain) | *Bhori* (a bird, Aldebaran, with two eggs) | *Medi* (part of a large farming scene also involving Canis Major, Lepus and Orion) |
| Crux | No mention | *Irukmara* or *Ipamaka* (a *Mahua* tree) | A *Mahua* tree. |
| Scorpius | No mention | *Borenagu* or Nago (a snake), or *Nagun* (a cobra); o *rTuntor* or *Tootera* (a scorpion) | *Michu* (responsible for producing the dead body, *Murda*, seen in Leo) |
| Cygnus | No mention | Cygnus = *Kavadi Kunde*, three pots stacked on top of each other. | No mention |
| Pegasus | *Samdur* (nearby animals predict the monsoon) | *Samdur* (nearby animals predict the monsoon) | No mention |
| Solar eclipses | Recognised but not explained | Indicates that payments are due to the Sun by humans or animals | No mention |

## 5 NOTES

1. In some regions of India occupied by Banjaras this ceremony is spelled and pronounced *Divali*.
2. Where alternative spellings of the same word are presented in Appendix 1, only the first version of the word, as it appears in the Appendix, is listed in this Table.

## 6 ACKNOWLEDGEMENT

We wish to thank the Raman Science Centre for providing us with the logistic support for this study. We also want to thank Professor Sir Arnold Wolfendale for his encouragement and support, and Professor Wayne Orchiston for all his efforts in making this paper more readable. Finally, we must also thank all our friends in the villages we visited for their enthusiastic response to our queries.

31, 95-102.

Stone, L., and Lurquin, P.F., 2010. *Genes, Culture and Human Evolution: A Synthesis*. Malden, Wiley-Blackwell.

Vahia, M.N., and Halkare, G., 2013. Aspects of Gond astronomy. *Journal of Astronomical History and Heritage*, 16, 29-44.

Vanjara, K.G., Naik, A., and Naik, I., 2012. *Banjara Logon ka Itihas*. Gandhinagar, Navdurga Publishers (in Hindi).

## APPENDIX 1: DETAILS OF THE VILLAGES VISITED DURING THE SURVEY

### A1.1 Banjara Villages Visited

Over a one week period we visited a total of 18 villages, 7 Banjara and 11 Kolam. The route taken by us is given in Figure A1.

#### A1.1.1 Dapora Village

Location 20° 12.27′ N, 78° 47.73′ E (225 meters above mean sea level).

Post: Dapora, Taluka Manora, District Washim. On the Karanja (Lad) – Manora Road, 16 km from Karanja.

Population: 8,000

Date of Visit: 2 May 2013.

Persons Met:
   Devananda Ratansing Rathod (M72)
   Puranasingh Mersing Rathod (M75)
   Shankar Janusing Rathod (M55)

Astronomical and Other Beliefs:
   Orion is *Harini* (a deer).
   *Jamakhat* (*Yamakhat*) – The first four stars of the Big Dipper form the cot used by *Yama*, the God of Death to ferry the dead. The remaining three stars form the funerary procession.

Spots on Moon appear to be an old lady sitting under a banyan tree spinning cotton yarn.

They consider the glow around the Moon important for predicting the monsoon.

*Porya Tara*: a morning star.

*Subtara*: evening star (star of good omen).

They know that there is a Pole Star.

Milky Way: *Mardaar wat*, the path of the dead where the person carried by *Jamkhat* eventually travels (presumably to heaven).

Taurus when seen in the east brings rain.

The Sun in called *Dan* (Din means day).

The Moon is called *Chanda* (similar to the Sanskrit name for the Moon).

Solar and lunar calendars are synchronised by using an intercalary month (*Dhonda Mahina*).

In Canis Major they can see the *Medi*, a pole where the bullock is tied for crushing the harvest to extract the grain.

They know that solar eclipses occur.

A comet is called a smoking star and is a good omen.

They know a meteor as *Tara tutgo*, a broken star.

They know the Pleiades as *Jhumko tara*, a decorative forehead element.

They sang a song saying "As the morning star rises, we begin loading grain … (on the back of a bull)."

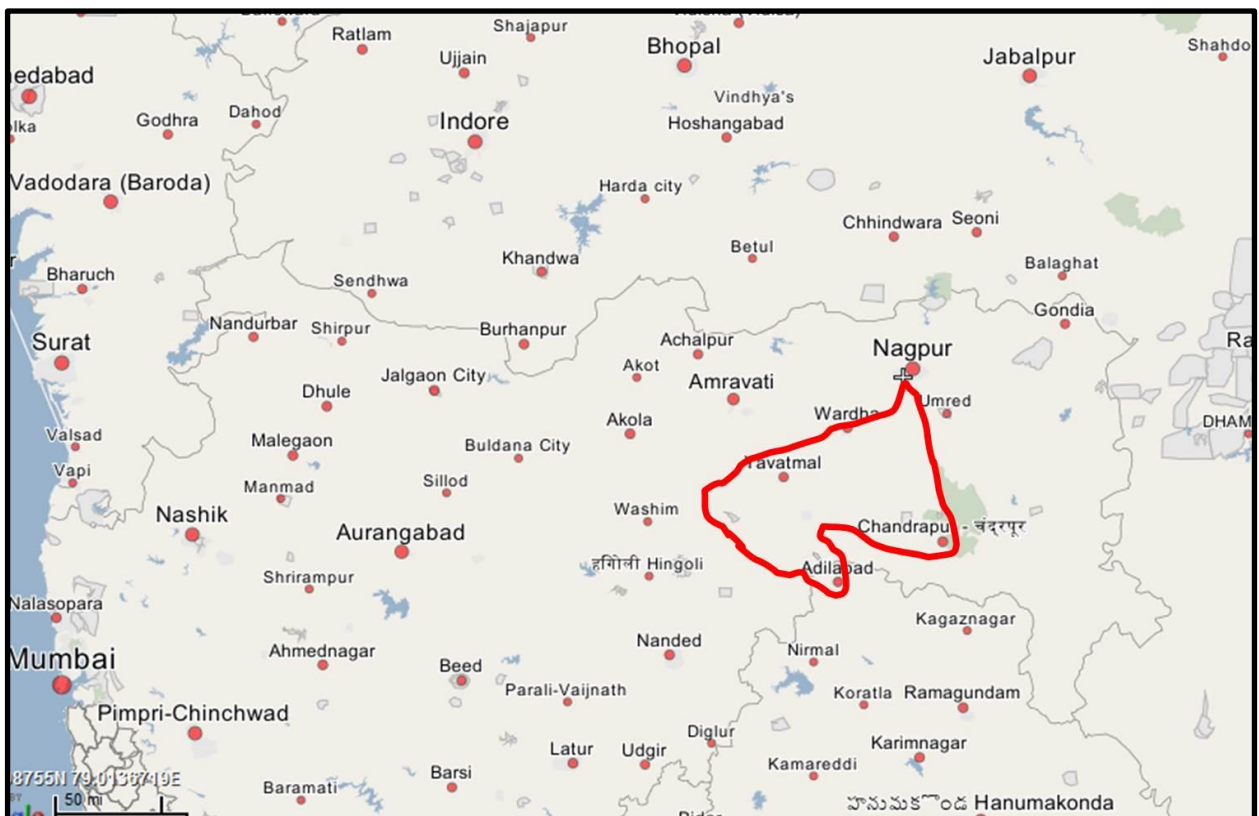

Figure A1: The route taken in the course of surveying Banjara and Kolam astronomy.

### A1.1.2 Phulumari Village

Location: 20° 5.759′ N, 77° 33.619′ E (415 meters above mean sea level).

Post: Phulumari, Taluka Manora, District Washim. On the Manora – Ratanwadi – Fulumari Road, 18 km from Manora. It also is approachable from Digras Pusad, Manora and Washim. It is surrounded by a hilly forest area. The place is a pilgrimage centre for Banjaras in association with Sewala Maharaj and a Garasha Temple of a Bull (*Nandi*).

Population: around 10,000. Claimed to be one of the biggest *Tanda* (Banjara villages) in India.

Date of Visit: 2 and 3 May 2013

Persons Met (brackets indicate subclan and age):
 Hirasing Nandu Rahod (Khola) (M80)
 Parasram Rama Rathod (Ransot) (M62)
 Bansi Shankar Rathod (Bhojawat) (M64)
 Babusing Bhola Rathod (Khetawat) (older than M55)
 Bhavurao Ganpat Chavan (Mood) (Sarpanch – village head) (M50)
 Pralhad Narsing Rathod (Khetawat) (M68)

Astronomical and Other Beliefs:
 *Porya Tara* is the Morning Star visible at 4 am (they have a song to celebrate the rise of this star that encourages them to begin their day). This indicates the last quarter of the night.
 Moonlight is called *Chandani* (a canonical word in Sanskrit and Marathi).
 The Pleiades are called *Shirser Jhumko* (an ornament for the forehead).
 Orion is *Halani*. Harvesting starts when Orion appears in the sky in the east and is completed before it begins to set at sunrise.
 The Big Dipper: *Jamkhat*, the bed of God *Yama* (god of death).
 The Moon as a forecaster of the monsoon – if there is an extended glow around the Moon then there will be good rains, but if it is restricted, the rains will be poor.
 Directional stars *Dhruv* (the Pole Star) shows north, *Vani tara* is a star at the zenith, and *Porya* (the morning star) shows east. North: *Dongari*; northeast: *Dharau* (and also *Ishan* in modern languages); east: *Suryatal* (*literally* the Sun's direction—the place from which the Sun (Surya) rises); southeast: *Agneya* (similar to *Agni* in modern languages); south: *Dakshin*, as in modern languages; as is *Pashim*: west.
 If it is cloudy to the east or there is a red glow at sunrise, it will rain heavily.
 If there is lightning in the southeast, rain is expected sometime between three hours and three days.
 If there is lightning in the northeast, there will be good rain.
 If a crow builds his nest at the top of a tree in May, less rain is expected, but if it makes it's nest lower good rains are expected from a direction such that the trunk of the tree will protect the nest.
 The Banjara remember having seen a comet (a star with a smoke) in 1965, most probably Comet Ikeya–Seki (C/1965 S1). It was seen for a fortnight until sunrise. In general a comet brings draught unless it comes from the north or the southeast, when it is a good omen. They also remember having seen a comet (a star with a tail) in March 1972 (the *Phalgun* month).
 A meteor is a broken star. A meteor that is seen to originate from the zenith and move to the east is a bad omen.
 If a male fox (*Salya*) barks at night it is a good omen, but if a female fox (*Sali*) barks it is a bad omen.
 They listed the seven important goddesses (and the animal they use for travelling).
 The dead are buried with the head to the west or the north.

### A1.1.3 Katarwadi Village

Location: 19° 56.52′ N, 77° 39.13′ E (348 meters above mean sea level).

Post: Kasola, Taluka Mahagaon, District Yavatmal. Twelve kilometres from Pusad, (a subdivisional headquarter) off the Pusad – Mahur Road.

Date of Visit: 3 May 2013.

Persons Met:
 L.D. Rathod (M58)
 Ajay Rathod (M40)
 Baburao Maharaj Khamalwadi (about M60)
 Baliram Maharaj Kalulalnagar Bori (about M60)
 Mohan Aade Manoharnagar Bori (about M60)
 Savairam Aade Maharaj Katarvadi (about M60)
 Pandurang Teja Rathod (about M65)
 Janusing Motiram Pawar Moli (ex Head) (about M50)
 Sajanibai Pandurang Rathod (F70)

Astronomical and Other Beliefs:
 The Morning Star: *Porya tara*.
 Orion is *Halani*. It is seen in the east and faces east-west and rises at 7 pm.
 *Saptarshi* is *Jamkhat* which keeps turning from east to west.
 The Pole Star is *Dhruv tara* (as in modern languages)
 The Milky Way is *Pandi*, the path of animals.
 The orientation of the Milky Way is north-south.
 In the Moon they can see a Banyan tree along with an old lady (*yadi*) spinning cotton yarn.
 Lightning in the north means rain.
 They know the relation between the glow around the Moon and the monsoon.
 The Evening Star is *Shubham* (an auspicious star).

They have seen a comet but do not know what it is.
A meteor brings death.

### A1.1.4 Kasola Village

Location: 19° 54.80′ N, 77° 39.7′ E (342 meters above mean sea level).
Post: Kasola, Taluka Mahagaon, District Yavatmal. Called *Sevalal Tanda*, it has about 900 houses and 3,755 cows. The main occupation used to be transport but is now farming.
Date of Visit: 3 May 2013.
Persons Met:
  Birbal Munna Chauhan (M88)
  Devlal Rodba Jadhav (M65)
  Haribhau Phulasingh Rathod (M70)
Astronomical and Other Beliefs:
  The Morning Star is *Pora* (*Porya*) *tara*.
  The Evening Star is *Sukachi Chandi* (Venus?), a glow that brings happiness. It is useful for telling time.
  Orion is *Halni*
  They know *Jamkhat*. They think that if is sinks the world will end (this is a partially borrowed idea from the Gonds).
  The Milky Way is the path of the dead (*mardarwat*).
  The Pleiades are remembered is the *Shirser Jhumko*.
  They know Taurus as *Kamedi* (the dove) which is close to *Halni* (Orion).
  They know that the Moon has dots, and a lunar halo indicates rain in slightly differently ways: if the glow is broken (*tala*) then the rains will be better, but if it is complete (*khala*) there will be little rain.
  Comets: *seshar tara*, and are auspicious.
  A meteor is a broken star and is a bad omen.
  They know about the crow's nest in May and its relation to the seasons.
  They know that lightning in the east brings rains in anywhere from 3 hours to 3 days.
  If there is a hailstorm in summer, the monsoon will be bad.
  If ants come out with eggs then the rain will be poor.
  They know that a rainbow (*Indra-dhanush* from modern languages) means the end of the current spell of rains.
  They have heard of twelve zodiacal signs but have no idea what they are.
  If a male fox (*Salya*) barks (*kolbhuki*) at night it is a good omen, but if a female fox (*Sali*) barks it is a bad omen.

### A1.1.5 Chinchkhed Village

Location 19° 52.25′ N, 78° 10.334′ E (270 meters above mean sea level).
Chinkhed Post: Chinchkhid Taluka Kinwat District: Nanded. On the Pusad – Mahur – Kinwat Road, 20 km from Mahur.
Date of Visit: 4 May 2013.
Persons Met:
  Bhimaniputra Mohan Naik (M66) [The author of several books and papers on the origin of the Banjaras.]
Astronomical and Other Beliefs:
  The Banjaras are the Pani tribe of Vedic literature and are snake worshipers. They are not Rajputs.
  Their main profession is transportation and trade—mainly in salt, cows, food grains and occasionally metals.
  Have often have been hounded as thieves.
  Their main identifier is the horned head gear.
  They worship *Shivaling* (the amorphous symbol of *Shiva*).
  The history of the Banjaras goes back to the Harappans. The Rajputs converted to the Banjaras *en mass* after the defeat of Rana Pratap against the fellow Rajputs in 1597 and reprisals were feared. Being providers of the Muslim army the Banjaras were exempt from persecution.
  They became farmers after the arrival of the railways made them redundant.

### A1.1.6 Palaitanda Village

Location 19° 37.776′ N, 78° 43.2′ E (305 meters above mean sea level).
Post Toyaguda, Mandal Bela, Block and District Adilabad. On the Adilabad – Satnala – Jamini – Palaitanda Road, about 26 km from Adilabad.
Date of Visit: 6 May 2013.
Persons Met:
  Chavan Dashrath (M65)
  Chavan Baliram (M70)
  Chavan Jogram (M60)
  Chavan Prakash (M50)
  Chavan Gunibai (F80)
  Chavan Hamalibai (F90)
  Rathod Shakuntala (F70)
Astronomical and Other Beliefs:
  Orion is *Halni*
  The Morning Star is *Porya tara*.
  They know that by using the Evening Star that rises in the east, one can determine the passage of time during the night.
  They know how to use a gnomon to determine the time of the day.
  They know that the Evening Star demands a sacrifice of a wheat bread (a *chapatti*)
  They know *Saptarshi* as *Jamkhat*.
  They know the Pleiades as *Jhumko*.

### A1.1.7 Nagrala Village

Location 19° 39.372′ N, 79° 6.604′ E (485 meters above mean sea level).
Post: Jiwti, Taluka Jiwti, District Chandrapur. On the Chandrapus – Rajura – Gadchandur – Bailampur – Nanakpathar – Nagarala Road,

16 km from Gadchandur and 70 km from Chandrapur.
Date of Visit: 8 May 2013.
Persons Met:
  Seva Amru Chauhan (M70)
  Vinayak Chandar Rathod (M46)
  Ramji Sakharam Rathod (M45)
  Motiram Todba Rathod (M65)
  Janabai Sakharam Jadhav (F60)
  Tayabai Chandar Rathod (F70)
  Malabai Motiram Rathod (F60)
  Sahebrao Sakharam Jadhav (M32)
Astronomical and Other Beliefs:
  A comet is a star with a tail and it brings a famine.
  The Big Dipper: (*Saptarshi*) *Jamkhat.*
  Orion is *Halni.*
  The Milky Way is the way to heaven.
  The Morning Star: *Porya Tara.*
  The Sun is *Dado* (not related to any Indian language), and the Moon is *Chanda* (similar to the modern Indian word) and they can see a tree on it.
  They know that eclipses occur but don't know why but they know that they should make donations at that time (a very Hindu concept).
  The Pleiades are *Jhumko*.
  Meteors are stars that come down by magic and are not good.
  They know the story of the crow and its nest.

**A1.2 Kolam Villages Visited**

A1.2.1 Dubaguda Village

Location: 19° 36.27′ N, 78° 45.086′ E (331 meters above mean sea level).
Post: Sayeedpur, Mandal Bela, District Adilabad, State Andhra Pradesh. Along the Satnala – Jamini – Sayeedpur – Dubaguda Road, 35 km from Adilabad.
The village is on the River Deyalamadg and is isolated from the main road by the River. There are about 80 houses. The population is about 500, 25 of whom are educated to 10th Standard (age 15 or so); there are 5 graduates. There is a primary school in the village.
Date of Visit: 5 May 2013.
People Met:
  Bapurao Tekam (M55)
  Lasama Tekam (M70)
  Maharu Tekam (M44)
  Lakuu Kodapa (M55)
  Lakshmibai Aatram (F52)
  Aaibai Tekam (F55)
  Bhimbai Kodapa (F47)
Astronomical and Other Beliefs:
  Followers of four gods by number–4th to 7th. (carried from old contact with the Gonds who have nine gods, number 1st to 9th; the Kolams were once priests of the Gonds).
  The counting system (for >3) is largely taken from Telugu.
  Directions are north: *Telganam*; south: *Kalam*; east: *Potkuranai*; west: *Potpodnavai.* East is also called *Suryatu* (the sunrise direction).
  They know of the Pleiades as *Kovela Kor*, a constellation with one large and several small birds. The Pleiades also are known as *Sappa*, an instrument that is used to beat the grain in order to separate the husk.
  Big Dipper (*Saptarshi*): *Mandater*, meaning a cot. The three following stars are three people, a Kolam, a Gond and a *Pardhan* (chief).
  They know Orion as *Tivpate*, an agricultural instrument used to sow the seeds.
  The Morning Star is *Vegud suka* and the Evening Star is *Jevan suka.*
  The Milky Way is *Margam,* a path.
  Taurus is *Bhori*, a bird (Aldebaran) with two eggs.
  Scorpius is *Tuntor* (a scorpion).
  Canis Major is *Mete*, the central pole around which a bull is tied and made to go around as it separates husk from grain. The ground is known as *Kalave.*
  They also see a group of three stars one on top of the other called *Kavadi Kunde* (Cygnus), representing three earthen-ware pots put on top of each other and seen in the northeast after sunset in May-June.

A1.2.2 Janguguda Village

Location: 19° 37.79′ N 78° 44.0′ E (318 meters above mean sea level).
Post: Saidpur, Mandal Bela, Block Adilabad, District Adilabad, State Andhra Pradesh. 30 km from Adilabad on the Adilabad – Satnala – Jamini – Palaitanda – Janguguda Road.
Date of Visit: 5 May 2013.
People Met:
  Sidam Surya (M60)
  Madavi Bapurao (M60)
  Sidam Bhimrao (M45)
  Madavi Kaniram (M28)
  Devbai Attram (F40)
  Jango Attram (M40)
Astronomical and Other Beliefs:
  The Big Dipper (*Saptarshi*) rises in the east in May. It is assumed to be a container with four edges whose opening is towards the zenith when it rises. In early June, at sunset it is close to zenith and 'upside down'. This is when it pours out the rains. The remaining three stars are on their way to drink the water.
  They know *Saptarshi* as *Mandter*, where the four main stars are a cot that is being sought by three people.
  The great square of Pegasus: *Samdur* (four stars), identified by the mention that it has four stars that appear near the zenith at sunrise in early May. If it appears bright the rains will be good.
  Five different animals appear near *Samdur*

(Pegasus). They are: *Namli* (a peacock), *Barre* (a buffalo), *Pande* (a frog), *Mekam* (a deer) and *Gurram* (a horse). The rain in a particular year depends on which animal is the most conspicuous in the sky at that time.

We suggest that the stars in the northeast in Pegasus is the deer; Cassiopeia is the peacock; Aquila is the horse; Cygnus is the buffalo; and Pisces is the frog.

The Pleiades is *Kovela Kor* (one large bird with lots of little birds). If it is sharp in the sky in June then the crop will be good.

Orion is *Tipan*, a set of three stars.

Sirius in Canis Major is *Mete*, the central pole around which a bull is tied and made to go around as it separates husk from grain. The ground is known as *Kalave*. *Adhara* in Canis Major is *Tiva Pate*, a three-legged stool on which one stands to drop the harvest onto the ground to separate husk from crop in the wind.

They know of the *Mahua* tree (*Ipamaka*) that is seen in October in the southern sky. This was identified as Crux by the Gonds.

They know of comets as *Sipursuka* (stars with tails). They consider them bad omens.

They refer to a meteor as *Suka erengten* (a broken star).

Scorpius is *Tuntoor* (a scorpion).

They refer to the Milky Way as *Panadi* or *Margam*, which means path. They know the Milky Way goes from northeast to southwest.

Leo is *Murda* (dead body) with mourners (*Ladvya*); this is almost a direct, but imperfect, memory of the Gond myth.

They know of eclipses. A total solar eclipse is good but a partial eclipse of the top of the Sun is bad for humans while a partial eclipse of the lower part of the Sun is bad for animals. The belief is that some tax collector has come to collect the taxes and if the Sun is fully engulfed the debt has been paid up in full, otherwise humans or animals have to pay the balance.

### A1.2.3  Jamini (Duttaguda) Village

Location: 19°39.105′ N, 78° 41.992′ E (302 meters above mean sea level).

Village Jamini (Dattaguda), Post Toyaguda, Mandal Jainad, Block and District Adilabad, State Andhra Pradesh. On the Adilabad – Satnala – Jamini – Dattagud Road, 25 km from Adilabad.

Date of Visit: 5 May 2013.

Persons Met:
  Mohan Pendur (Village Head) (M42)
  Bapurao Kodapa (M42)
  Mesram Shankar (M53)

Astronomical and Other Beliefs
  The Big Dipper is *Mandter*.
  The Pleiades is *Kovela Kor* (one big and lots of small birds).
  The Morning Star is *Vegud Suka*.

The Evening Star is *Jevan* (dinner) *suka* (star).

They know *Irukmara* as the tree of *Mahua* in Crux. They know the neighbouring small stars as flowers of *Mahua* (*Irup pokke)*. One bright star (α Centauri) in the region is called *Murta* (an old lady) and β Centauri is *Kadma* (a young lady). They are there to collect the flowers.

They know a constellation called *Kawadi Kunde* (Cygnus) which looks like three pots on top of each other. This constellation rises in the northeast in May-June early in the evening.

They know an asterism called *Dandare Pila* which is an anticlockwise spiral that rises in October-November in the northeast. If it is bright and clear, it is good omen, but if it is hazy this is a bad omen. This may be the tail of Scorpius.

They know of the Milky Way as 'The Path', or the animal path.

They know of Taurus as a bird with eggs.

They also know Scorpius as *Borenagu* (a snake), which is seen at sunrise in *Pusa maas* (January).

They identify Sirius in Canis Major that rises January (*Pusa maas*) as *Mete*, the central pole around which a bull is tied and made to go around as it separates husk from grain. The ground is known as *Kalave.* *Adhara* in Canis Major is *Tiva Pate*, a three-legged stool on which one stands to drop the harvest into the ground to separate the husk from the crop in the wind.

Orion: *TIpun*.

They know of meteors and *Suka Yerangtin* (a falling star), and recall that it once fell on a tree and burning it down.

A comet is *Sipur suka* (a star with a tail), and is considered a bad omen.

They know the solar eclipse story.

### A1.2.4  Laxmipur Village

Location: 19° 41.809′ N, 78° 21.049′ E (345 meters above mean sea level).

Post: Kuchalapur, governed from Kochlapur, Mandal Talamadgu, Taluka and District Adilabad. On the Mahur – Adilabad Road, 25 km before Adilabad. The first village in Andhra Pradesh on the road from Mahur.

The village has 48 houses and a population of about 200.

Date of Visit: 6 May 2013.

Persons Met:
  Kadopa Bhimrao (M45)
  Kumra Bhimrao (M55)
  Tekam Kappu (M60)
  Aatram Tukaram (M45)
  Tekam Laxmibai (F70)
  Tekam Bhimbai (F65)

Astronomical and Other Beliefs:
  The Morning Star is *Vegud Suka.*

The Evening Star is *Jevan Suka* (dinner star).
The Pleiades is *Kovla Kore.*
The Big Dipper is *Mandetar*, a cot being pursued by the people, Kolvar, Gond and Patadi.
The Milky Way is *Panadi*.
Pegasus is *Samdur*.
They are clear about Orion, including its entire structure, not just the belt.
Scorpio is *Nagun* (the cobra), that rises at 4 am.
Cygnus is *Kavdi Kunde.*
They know of Crux as the *Mahua* tree.
They think of meteors as stellar excreta.
Their month goes from New Moon to New Moon.
They do not know the arrangement around Sirius.
They know a rainbow as *Ayak*, which will end the rain.
They know the complete story of solar eclipses.

### A1.2.5 Sonapur Village

Location: 19° 41.486′ N, 78° 21.905′ E (336 meters above mean sea level).

Village Sonapur, Post Kuchalapur, Mandal Talamadgu, Block and District Adilabad. On the Adilabad – Sunkadi – Lingin – Sonapur Road, about 20 km from Adilabad.

Date of Visit: 6 May 2013.
People Met:
　Tekam Letanna (M60)
　Tekam Rambai (MF60)
　Tekam Muttubai (M55)
　Aatram Aiyu (M55)
　Madavi Supari (M35)

Astronomical and Other Beliefs:
　The Morning Star is *Vegud Sukum*.
　The Evening Star is *Jevan Sukum*.
　Pegasus is *Samdur*, including five animals around it: *Namli* (a peacock), which is not good; *Barre* (a buffallo), which is excellent; *Pande* (a frog), which brings good rains; *Mekam* (a deer), which brings average rains and *Gurram* (a horse) which brings average rains.
　They know of the cot in the Big Dipper (*Saptarshi*), with a Kolam, a Gond and a Patadi in pursuit of it.
　The Milky Way is *Bhori Sanko*, an animal path.
　Cygnus is *Kavedi koda*.
　Scorpius is *Borenag* (the cobra), which rises in the northeast, south of the zenith seen in *Pusa Maas* (February).
　They know the story of eclipses and identify three different types depending upon whether the upper, lower or middle part of the Sun is covered. The first is bad for humans, the second is bad for animals, and the third is good for humans and animals.
　They know about the glow around the Moon: when the aura is small the rains are far but if the glow is large the rains are nearby.
　The Pleiades is *Kovela Kore*.
　Orion is *Tipan*.
　The region around Sirius is known as *Met*.
　They know of Crux as *Ipamak*, the *Mahua* tree.

### A1.2.6 Landgi Pod Village

Location: 19° 52.522′ N, 78° 39.753′ E (273 meters above mean sea level).

Post: Matharjun, Taluka Zarijamni, District Yeotmal, State of Maharashtra. On the Bori – Mandavi – Piwardol – Gavara – Landgi Pod Road, about 10 km from Bori and 2 km from Matharjun.

Date of Visit: 7 May 2013.
Persons Met:
　Bajirao Punjaram Madavi (M60)
　Laxman Punjaram Madavi (M55)
　Kishore Bhovanu Mesram (M21)
　Mahadev Punjaram Madavi (M65)

Astronomical and Other Knowledge:
　The Morning Star is *Vegun suka.*
　The Evening Star is *Jevean Suka.*
　The Pleiades is *Kovala Kor.*
　The Belt of Orion is *Kavadi Kunde.*
　The cot of the Big Dipper is being stolen by three thieves who are following it (this story is taken from the Gonds).
　Orion is known as *Tipan.*
　They know the region around Orion as a farming scene with *Medi, Kalave* and *Tandel/ Aidal* (a bull).
　The great square of Pegasus is *Samdur*, that rises at 3-4 am. They know that if the pig (*Ture*) is close to *Samdur*, it will be a bad monsoon; if it is *Mekam* (*Rohi*) or the deer (*Gorya*) it will be average; if it is the buffalo (*Sir*) it will be an excellent monsoon; if it is *Pande* (*Beduk*, the frog) it will be a good monsoon.
　They know that monsoon arrives in the month of *Budbhavai* (June).
　They know Crux as *Irukmara* (*Ipemak*), that is the tree of *Mahua*. Their story of Crux is slightly different, with Alpha and Beta Centauri called *Oedda* (man) and *Pilla* (woman), instead of old and young lady, respectively.
　Scorpius is *Nagpan* (the cobra).
　A comet is *Sipharsukka* (a star with a tail).
　They know the story of solar eclipses, but in reverse: if the bottom of the Sun is covered it is a bad omen for humans, but if the upper half is covered it is bad for animals.
　They know of Taurus as the constellation of one bird with two small eggs (*Bhori*).
　About lunar eclipses, the eclipse is dark if a frog (*Beduk*) attacks the Moon and pale if a caterpillar (*Gongpatad*) is eating the Moon.
　A rainbow is the *Bhimayak's* bow.

### A1.2.7 Chinchpod Village

Location: 19° 54.589′ N, 78° 40.432′ E (283 meters above mean sea level).

Post: Matharjun, Tahsil Zarijamni, District Yeotmal. On the Pkandarkawa – Shibla – Matharjun Roa,. 28 km from Pandharkawada.
Date of Visit: 7 May 2013.
Persons Met:
  Anandrao Tekam (M55)
  Layanibai Tekam (F80)
Astronomical and Other Beliefs:
  The Evening Star is *Jevan Suka.*
  The Morning Star is *Vegud Suka.*
  The Pleiades is *Kovela Kore.*
  Cygnus is *Kawadi Kundel.*
  They know of the cot in the Big Dipper (*Ter*) and the thieves (*Dongal*).
  Pegasus is *Samdur*. *Namli*: more rain; Peacock: less rain; *Mekam*, *Rohi*: more rain; *Ture*, *dukkar* (the pig): a good monsoon is expected; *Gorya* (the deer): less than normal rain; *Pande*, *Beduk* (the frog): an excellent monsoon is expected.
  They know of Taurus is a bird with eggs.
  They know of Scorpius as a cobra (*Bornagu* or *Nagpan*).
  Orion is *Tipan*.
  They know the story of solar eclipses.
  Burials are east-west, with the head facing the west.

### A1.2.8  Bhimnala Village

Location: 19° 55.603′ N, 78° 40.162′ E (269 meters above mean see level).
Post: Matharjun, Taluka Zarijamni, District Yeotmal Stat Maharashtra. On the Shibla – Rajni – Shiratoki – Bhimnala Road, 6 km from Shibla.
Date of Visit: 7 May 2013.
Persons Met:
  Bhima Tukaram Aatram (M65)
  Sonu Letu Tekam (M35)
  Lalu Surya Tekam (M60)
  Dilip Bhima Tekam (M22)
  Sudhakar Aayya Aatram (M24)
Astronomical and Other Beliefs:
  The Pleiades is *Kovela Kora*, a group of birds.
  Cygnus is *Kuvadi Kunde.*
  They know the Morning Star as *Pele suka.* However, the star for waking up is *Veka Potur* (the son of the morning).
  The Evening Star is *Jevan Suka.*
  They know *Samdur*, including the relation between the intensity of rains and the animals (*Guram*, the horse) etc.
  They know Crux as the *Mahua* tree (*Irupmak*).
  They know Scorpio as a cobra.
  A comet as a broom star.
  They know about solar eclipses.
  For lunar eclipses, they suggest that if the Moon is being eaten by a scorpion it will be red but if a frog is eating it, it will be black.

### A1.2.9  Raipur (Khadki) Village

Location: 19° 39.904′ N, 79° 5.957′ E (508 meters above mean sea level).
Post: Jiwati, Taluka Jiwati District Chandrapur. On the Chandrapur – Rajura – Ghadchandur – Bailampur – Nanakpathar – Nagrala Road, 3 km from Nagrala.
Date of Visit: 8 May 2013.
Persons Met:
  Soma Rama Sidam (M60)
  Jhadu Boju Kodape (M50)
  Jaitu Bhima Sidam (M42)
  Maru Jangu Kodape (M45)
  Ayyo Tukaram Sidam (M35)
Astronomical and Other Knowledge:
  Farming is by tilling the land, by hammering to break the ground, and then sowing.
  They know of *Samdur* and the animals that approach it decide on the rain: *Turre*, *Dukkar* (the pig) good rain; *Mokam* (the deer), *Gurram* and *Tumali*: average rain; *Barre*, *mais* (the bullock): lots of rain.
  *Saptarshi* is a cot, *ava ter,* which is being pursued by three thieves, *Dongal* a Kolam, a Gond and a Patadi.
  The Morning Star is *Vegur Suka.*
  The Evening Star is *Jevan Suka.*
  The Pleiades is *Kovela Kor.*
  Unlike other villages, they state that *Kavde Kunde* is the belt of Orion and *Tipan* is the sword of Orion (in other villages *Kavde Kunde* is Cygnus).
  They have heard of Leo (as described by the Gonds).
  Scorpius is *Tootera* (a scorpion), but it also is referred to as a snake (*nago*).
  They know of a counter-clockwise spiral of stars that also looks like a snake about 10° east of Scorpius.
  The Milky Way is the path of animals.
  They know stories of solar and lunar eclipses.
  They also know Taurus as a bird with two eggs.
  Crux is *Ipemak.*
  They know *Tiva*, *Miti*, *Kovela* and *Konda* around Canis Major. These are various tools associated with farming and are described in detail by the Gonds.
  They know a comet as *Sipur Suka* (a star with a tail), and consider it a bad omen.
  A rainbow is *Aikawa.*
  Burials are oriented east-west, with the head to the east. The burial site is away from the village. A commemorative community site as per the god is created in the village and is worshipped as a common site for all ancestors on important occasions.

### A1.2.10 Kakban Village

Location: 19° 38.743′ N, 79° 7.038′ E (487 meters above mean sea level).
Post: Jiwati, Taluka Jiwati, District Chandrapur about 70 Kolam houses with very few Gonds

and Banjaras.
Date of Visit: 8 May 2013.
Persons Met:
  Nanaju Maru Madavi (M43)
  Bhima Raju Aatram (M40)
  Raju Mutta Aatram (M50)
  Bhujangrao Marotrao Kotnake (M50)
  Netu Naru Sidam (M50)
  Maru Jangu Kodape (M40)
Astronomical and Other Knowledge:
  They know of *Samdur,* with *Namali* bringing less rain, *Turre* and *Barre* bringing moderate rain, *Gurram* bringing less of rain and *Pande* bringing good rain.
  The Pleiades: is *Kovela Kor*, with the little bright star at the base called *Kovela Komdi*. At maximum it rises 70° above the horizon.
  The belt of Orion is *Kavadi Kunde*, and the sword is *Tipun*.
  *Saptarshi* is a golden cot that is being pursued by a Kolam, a Gond and a Patadi, in that order.
  Crux is the *Mahua* tree, with the bright star near it an old lady and the dimmer star a young girl.
  The Evening Star is *Jevan Suka.*
  The Morning Star is *Vegud Suka.*
  A meteor is *Suka Erangtu*, and is a bad omen.
  They know Canis Major as *Medhi*, *Khala*, and *Banati*, *Tlva pate*. These are images from the Gonds, who view the entire region as a farming scene.
  A comet is *Sipur Suka*, and is a bad omen.
  Scorpius is *Tuntoora*, the scorpion.
  The Milky Way is *Jeev* (a path, the path of life).
  They know the complete story of solar eclipses.
  They know the story of lunar eclipses. If the frog is attacking the Moon the eclipse will be dark red, and if it is being eaten by a caterpillar the eclipse will be lighter.
  A rainbow is *Ayyakawa*, which belongs to the grandparents (the grandfather, *Ayyak*, towards the violet and the grandmother, *awa*, towards the red).
  Cardinal directions are: *Kaland* (north), *Potkurina* (east), *Telagnaam* (south) and *Potpadna* (west).
  Burials are oriented east-west with the head to the east. They worship the ancestors. A memorial pillar if made from stone is called *Gunda* and if made of wood is called *Mundem*.
  A month is from New Moon to New Moon, and one month is added every three years.
  They know of the glow around the Moon and the rains.

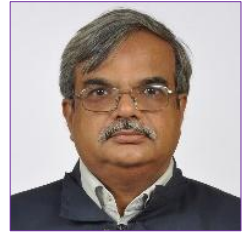

Mayank Vahia is a Professor in the Department of Astronomy and Astrophysics at the Tata Institute of Fundamental Research in Mumbai. He has a wide range of interests in the origin and the growth of astronomy in India. Mayank is a member of IAU Commissions 41 (History of Astronomy) and 44 (Space & High Energy Astrophysics), and has published about 30 papers on the history of astronomy.

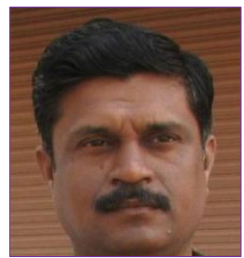

Ganesh Halkare is is an advocate in Amravati, a town in Maharashtra. He also has a post-graduate degree in Archaeology and Anthropology from Nagpur University. He has a deep interest in tribal education, particularly in the removal of superstition among tribe members. He also is deeply interested in tribal anthropology and is highly respected amongst the tribesmen for his work in ensuring that they are aware of and can exercise their rights within the nation state. Ganesh has published more than a dozen research papers on the archaeology of the Nagpur region. He is now working on the astronomy of other tribes in the Nagpur region.

Kishore Menon is Public relation officer at the Tata Institute of Fundamental Research. He has done a certificate course in Journalism. He has a deep interest in the nature of Indian civilisation and different aspects of Indian culture.

Harini Calamur is a media professional specialising in educational media and social media. She is also an avid student of Indian culture, and her passion is the use of social media for communication and development.                                                                                 .